\documentclass[aps,prx,twocolumn, superscriptaddress,amsmath,amssymb,longbibliography]{revtex4-2}
\usepackage{ulem}
\usepackage{float}
\usepackage[english]{babel}
\usepackage[utf8]{inputenc}
\usepackage[colorinlistoftodos, color=green!40, prependcaption]{todonotes}

\begin{document}
\preprint{APS/123-QED}
\title{Fragility of topology under electronic correlations in iron chalcogenides}

\author{Younsik Kim}
\email[]{leblang@snu.ac.kr}
    \affiliation{Department of Physics and Astronomy, Seoul National University, Seoul, Korea}
\author{Junseo Yoo}
    \affiliation{Department of Physics and Astronomy, Seoul National University, Seoul, Korea}
\author{Sehoon Kim}
    \affiliation{Department of Physics and Astronomy, Seoul National University, Seoul, Korea}
\author{Sungsoo Hahn}
    \affiliation{MAX IV laboratory, Lund University, Lund, Sweden}
\author{Kiyohisa Tanaka}
    \affiliation{National Institutes of Natural Science, Institute for Molecular Science, Okazaki, Japan}
\author{Li Yu}
    \affiliation{Beijing National Laboratory for Condensed Matter Physics and Institute of Physics, Chinese Academy of Sciences, Beijing 100190, P. R. China}
    \affiliation{University of Chinese Academy of Sciences, Beijing 100049, P. R. China}
\author{Minjae Kim}
    \email[]{mjkim1985@jbnu.ac.kr}
    \affiliation{Korea Institute for Advanced Study, Seoul, Korea}
    \affiliation{Department of Semiconductor Science and Technology, Jeonbuk National University, Jeonju, Korea}
\author{Changyoung Kim}
     \email[]{changyoung@snu.ac.kr}
    \affiliation{Department of Physics and Astronomy, Seoul National University, Seoul, Korea}
\date{\today} 

\begin{abstract}
The interplay between electronic correlations and topology is a central topic in the study of quantum materials. In this work, we investigate the impact of the orbital-selective Mott phase (OSMP) on the topological properties of FeTe$_{1-x}$Se$_x$ (FTS), an iron chalcogenide superconductor known to host both non-trivial Z$_2$ topology and strong electronic correlations. Using angle-resolved photoemission spectroscopy, we track the evolution of topological surface states across various doping levels and temperatures. We identify a topological phase transition between trivial and non-trivial topology as a function of selenium content, with critical behavior observed between $x$ = 0.04 and $x$ = 0.09. Additionally, we find that at elevated temperatures, the coherence of the topological surface state deteriorates due to the emergence of OSMP, despite the topological invariant remaining intact. Our results demonstrate that the non-trivial topology in iron chalcogenide is fragile under strong electronic correlations.
\end{abstract}


\maketitle

\section{Introduction} \label{sec:outline}

With the discovery of topological phases in condensed matter, intriguing and unprecedented phenomena have been reported in materials with non-trivial topological invariants. A notable aspect of the non-trivial topology is that topological properties can be effectively coupled with various degrees of freedom therein, allowing extended functionalities~\cite{qi2011topological,hasan2010colloquium}. A representative example of this is topological superconductivity induced by the co-emergence of the non-trivial $Z_2$ topology and superconductivity in FeTe$_{1-x}$Se$_x$ (FTS)~\cite{zhang2018observation, zhang2019multiple}. In this context, it has been extensively studied how topological phases are coupled with other degrees of freedom.

\begin{figure}[!t]
	\centering
	\includegraphics[width=0.5\textwidth]{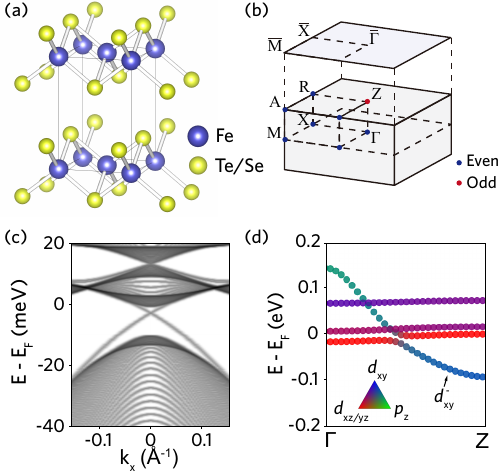}
	\caption{
         {\bf Crystal and electronic structure of FeTe$_{1-x}$Se$_x$ (FTS)}
         (a) Crystal structure of FTS.
         (b) Schematic of the Brillouin zone of FTS. Blue and red dots at the time-reversal invariant momentum points indicate the even and odd parities of the bands, respectively.
	(c) Linearized quasiparticle self-consistent GW plus dynamical mean-field theory plus spin-orbit coupling (LQSGW+DMFT+SOC) calculation of the projected band structure onto surface Brillouin zone along $\overline{\Gamma}-\overline{\mathrm{X}}$.
     (d) LQSGW+DMFT+SOC calculation of the band structure along $\Gamma-\mathrm{Z}$. Blue, red, and green colors correspond to the projected orbital character of $d_{xy}$, $d_{xz/yz}$, and $p_z$, respectively. The weight of the $p_z$ orbital is multiplied by a factor of 4 for better visualization.
	} 
	\label{Fig1}
\end{figure}
\begin{figure*}[ht!]
	\centering
	\includegraphics[width=1\textwidth]{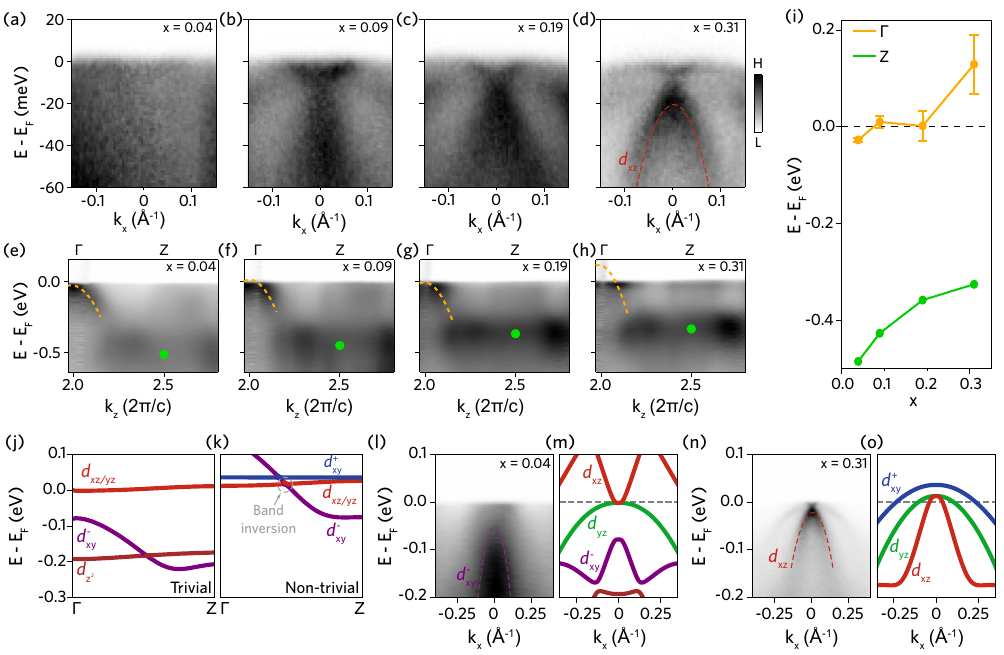}
	\caption{
         {\bf Doping dependence of the electronic structures of FTS.}
        (a-d) Band dispersion along $\overline{\Gamma}-\overline{\mathrm{X}}$, recorded with a $p$-polarized 7~eV light at a temperature of 15 K, with selenium contents of $x$ = 0.04, 0.09, 0.19, and 0.31, respectively.
        (e-h) Band dispersion along $\Gamma-\mathrm{Z}$ at 15 K, with selenium contents of $x$ = 0.04, 0.09, 0.19, and 0.31, respectively. The yellow dashed lines and green dots represent the fitted peak positions of the $d_{xy}^-$ band near the $\Gamma$ point and at the Z point, respectively.
        (i) Energy of the odd-parity $d_{xy}^-$ band at the $\Gamma$ point (yellow line) and $Z$ point (green line) as a function of selenium content $x$. Error bars represent the standard deviation obtained from multiple fittings using different random seeds.
        (j,k) LQSGW+DMFT calculation along $\Gamma-\mathrm{Z}$ with anion heights of $Z_{Se}=1.60~\mathrm{\AA}, 1.48~\mathrm{\AA}$, respectively.
        (l,n) Band dispersion along $\Gamma-\mathrm{M}$ at $k_z \approx 0$ with $x = 0.04$ and $x = 0.31$, respectively.
        (m,o) Corresponding LQSGW+DMFT calculation with anion heights of $Z_{Se}=1.60~\mathrm{\AA}, 1.48~\mathrm{\AA}$, respectively.     
        }
	\label{Fig2}
\end{figure*}

On the other hand, emergent phases in quantum materials typically involve electronic correlation. In unconventional superconductors, for instance, electronic correlations are crucial ingredients for the emergence of superconductivity~\cite{paglione2010high,fernandes2014drives}. However, topology has been mainly discussed within the concept of an effective single-particle scheme, where electron quasiparticles are independent fermions. This scheme is no longer an appropriate description when many-body electronic correlation effects set in~\cite{hasan2010colloquium,rachel2018interacting}. An extreme case is the Mott insulating phase from strong electronic correlations, in which the associated electron is spatially localized, resulting in a diverging incoherence of the electron's quasiparticle with a diminishing residue. Therefore, investigating how topological invariants evolve as a function of the quasiparticle residue can provide new insights into the interplay between topology and such correlation effects. It is noteworthy that SmB$_6$ is a canonical system exhibiting non-trivial topology in a strongly correlated regime~\cite{li2020emergent,neupane2013surface,jiang2013observation}. However, detailed investigation of the topological surface state has been hindered by the limited accessibility of laser-based ARPES and the surface state's sensitivity to impurity levels~\cite{jiao2018magnetic,hlawenka2018samarium}.

FTS represents an ideal platform for examining this interplay, as it simultaneously hosts non-trivial $Z_2$ topology and an orbital-selective Mott phase (OSMP) as a function of Se contents~\cite{checkelsky2024flat,li2024orbital}. The OSMP is a unique state where one band becomes localized while others remain itinerant~\cite{yi2017role,yin2011kinetic}. By controlling temperature and chemical doping, the quasiparticle residue of the localized orbital can be effectively tuned in the vicinity of the OSMP, thereby modulating the topological properties. Employing angle-resolved photoemission spectroscopy (ARPES), we could directly observe and map this topological phase transition. Specifically, we observed a topological phase transition at the lightly Se-doped regime in FTS, resulting in trivial topology at the Te end. Temperature-dependent measurements reveal no significant band renormalization, which conserves the topological invariant. However, the topological surface state becomes rapidly incoherent at elevated temperatures, which is an indication of broken topological properties. By firstly observing the doping- and temperature-dependent evolution of the $k_z$ dispersion of the band associated with topology, our study provides crucial insights into the interplay between topology and electronic correlations.

\begin{figure*}[hbt!]
	\centering
	\includegraphics[width=1\textwidth]{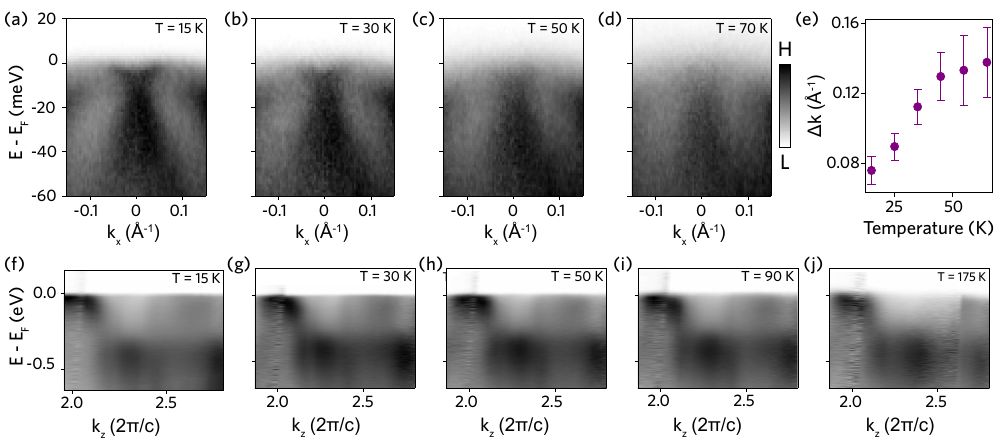}
	\caption{
        {\bf Temperature dependence of the electronic structures of FTS.}
        (a-d) Band dispersion along $\overline{\Gamma}-\overline{\mathrm{X}}$, recorded with $p$-polarized 7~eV light at temperatures of 15, 30, 50, 70 K, with a selenium content of $x$ = 0.19, respectively.
        (e) Full width at half maximum of the Dirac surface state extracted from momentum distribution curve fittings. The error bars represent the standard deviation of the fitted parameter. Note that the error bars for the 15 K and 30 K data points are smaller than the size of the symbols. 
        (f-j) Band dispersion along $\Gamma-\mathrm{Z}$ with selenium contents of $x$ = 0.09 at temperatures of 15, 30, 50, 70, and 175 K, respectively.
	} 
	\label{Fig3}
\end{figure*}

\section{Results} \label{sec:Results and Discussion}

FTS has the simplest crystal structure among iron-based superconductors (Fig. 1(a)), yet it exhibits diverse phenomena as a function of chalcogen contents, as shown in the sensitively controlled correlated electronic structures~\cite{yi2017role}. Non-trivial Z$_2$ topology is a prime example; it arises from a band inversion along the $\Gamma - Z$ direction, leading to an odd-parity occupied band at the Z point, while bands at other time-reversal invariant momenta remain even parity (see Fig. 1(b))~\cite{wang2015topological,zhang2018observation,kim2024orbital}. As a result of the band inversion, a Dirac surface state is formed at the $\Gamma$ point (Fig. 1(c)). This band inversion is associated with the crossing of the odd-parity $d_{xy}^-$ and even-parity $d_{xz}$ bands near the Fermi level (Fig. 1(d)). The odd-parity band is a hybridized state with $d_{xy}^-$ and $p_z$ orbitals~\cite{kim2024orbital}. We hereafter define the odd-parity band as the $d_{xy}^-$ band. Due to the significant interlayer overlap of $p_z$ orbitals, the odd-parity band is highly dispersive along the $k_z$ direction as illustrated in Fig. 1(d). This significant overlap and hybridization make the position of the $d_{xy}^-$ band
sensitive to the Se/Te ratio, which allows us to control the topological properties~\cite{wang2015topological,kim2024orbital}. With its highly tunable topology and the presence of an OSMP, FTS provides an ideal platform to explore the interplay between electronic correlations and topology.

To reveal the topological properties of iron chalcogenide superconductors in the strongly correlated limit, we investigated the Dirac surface state using high-resolution laser-based ARPES (see Supplemental Material (SM) for the experimental geometry and photoemission selection rule~\cite{supple}). Figures 2(a-d) show the laser-based ARPES results near the $\Gamma$ point obtained with selenium contents $x$ of 0.04, 0.09, 0.19, and 0.31, respectively. The ARPES results reveal that the dopings of $x$ = 0.09, 0.19, and 0.31 show the Dirac surface state (Fig. 2(b-d)), while the doping of $x$ = 0.04 does not show the Dirac surface state (Fig. 2(a)). We note that the selenium content $x$ was determined using energy dispersive spectroscopy (see SM for sample characterizations~\cite{supple}).

To further investigate the origin of the missing Dirac cone, we performed ARPES measurements along the $k_z$ direction on FTS with the same doping series. Shown in Fig. 2(e-h) are the electronic structures of FTS along the $\Gamma-Z$ line with selenium contents of $x$ = 0.04, 0.09, 0.19, and 0.31, respectively. They all show a dispersive band, which can be attributed to the $d_{xy}^-$ band. While the $d_{xy}$ band typically exhibits weak spectral weight, the unexpectedly strong intensity observed here is attributed to the $d_{xy}^-$ state’s strong hybridization with $p_z$ orbitals, which serves as the main contributor to the spectral weight. The seemingly discontinuous band is further attributed to matrix element effects (see SM for detailed discussions~\cite{supple}). To track the evolution of the $d_{xy}^-$ band as a function of doping, the top and bottom positions of the band at the $\Gamma$ and $Z$ points, respectively, are extracted from two-dimensional fittings as shown in Fig. 2(i) (see SM for details of the fitting~\cite{supple})~\cite{kurleto2021two}. The $d_{xy}^-$ band overall moves towards lower energy for lower Se contents. This doping-dependent evolution of the $p_z$ band is consistent with previous calculations on FeTe$_{0.55}$Se$_{0.45}$ and FeSe~\cite{wang2015topological}. The notable aspect of the doping-dependent evolution is that the band at the top of the $d_{xy}^-$ band at the $\Gamma$ point goes below the Fermi level at $x$ = 0.04. If the $d_{xy}^-$ band goes below the $d_{xz}$ band, which is near the Fermi level, this makes the inversion parities of the $\Gamma$ and $Z$ both odd, resulting in trivial topology. This feature is indeed observed in the calculated band structures, illustrating the evolution of the $d_{xy}^-$ band (see Fig.~2(j-k)).

Further evidence for the change in the band inversion condition is provided by the in-plane band structures. Our linearized quasiparticle self-consistent GW plus dynamical mean-field theory (LQSGW+DMFT) calculations shown in Fig.~2(m,o) predict that upon the topological phase transition, the $d_{xz}$ band is no longer a hole-like band near the $\Gamma$ point at the Fermi level. Instead, the hole-like $d_{xy}^-$ band occupies a higher binding energy. Consequently, the absence of the hole-like band near the Fermi level serves as a proxy for the topologically trivial state. Consistent with the calculations, the data for $x = 0.04$ shows a hole band with the band top around – 50~meV (Fig.~2(l)), while the data for $x = 0.31$ shows a well-defined $d_{xz}$ band as well as the Dirac cone (Fig.~2(n)). These observations are also consistent with our laser-ARPES measurements shown in Fig. 2(a-d), which exhibit the absence of the $d_{xz}$ band and Dirac cone for $x = 0.04$. Taking into account the ARPES results and calculations shown in Fig.~2, we conclude that the quantum topological phase transition occurs between $x$ = 0.04 and 0.09, resulting in a trivial topology in $x$ = 0.04.

Another notable aspect of the evolution of the $d_{xy}^-$ band is that the overall energy dispersion, determined by the energy difference between the $\Gamma$ and Z points, does not change considerably as a function of doping. This behavior cannot be solely explained without correlation effects, as a previous study based on the density functional theory predicted that the bandwidth of $d_{xy}^-$ strongly increases with Te doping since Te doping enhances the interlayer coupling from the out-of-plane tunneling of the chalcogen $p_z$ orbital~\cite{wang2015topological}. In addition to the aforementioned non-interacting term, the correlation effect renormalizes the $d_{xy}^{-} - p_{z}$ hybridization by a factor of $\sqrt{Z_{xy}}$ where $Z_{xy}$ is the quasiparticle residue of $d_{xy}$ orbital, reduced from the enhancement of the Te ratio or the anion heights by approaching the OSMP~\cite{kim2024orbital}. These two opposite terms compensate for each other while $1/Z_{xy}$ (the effective mass of $d_{xy}$) does not diverge at low temperatures, resulting in an unchanged bandwidth.

Having established the doping dependence of topology at the lightly Se-doped limit, we now investigate the temperature dependence of the Dirac surface state on the compound with $x$ = 0.19, which exhibits a clear Dirac surface state at low temperatures. Fig. 3(a-d) illustrate the Dirac surface state with temperatures of 15 K, 30 K, 50 K, and 70 K, respectively. The Dirac surface state is well-defined at 15 K. The Dirac surface state in this compound becomes rapidly broadened at relatively low temperatures, as evidenced by the loss of spectral weight and extracted linewidth. The Dirac surface state is already nearly indistinguishable at 50~K.

To quantitatively investigate the temperature dependence of the Dirac surface state, we performed momentum distribution curve fitting of the Dirac surface state at the Fermi level and plotted the full width at half maximum as a function of temperature in Fig. 3(e) (see Methods for fitting details). An abrupt change in the mean free path is observed between 30 K and 50 K, as also seen from the blurred spectrum at 50 K. Above 50 K, the mean free path estimated from the momentum width is less than $7~\mathrm{\mathring{A}}$, which is strikingly short for a topologically protected surface state. Given that typical topological surface states exhibit a mean free path on the order of tens of nanometers at room temperature~\cite{tian2013dual,pan2013persistent,pan2013persistent}, the unusually short mean free path in this compound at relatively low temperatures suggests the presence of additional modulation that scatters the topological surface state.

One possible scenario for the loss of the Dirac surface state is a transition to trivial topology at high temperatures. A previous study based on dynamic mean-field theory calculations predicted that OSMP could alter the overall band structures suppressing the band inversion, as the $d_{xy}^-$ band disappears near the Fermi level in the OSMP~\cite{kim2024orbital}. To confirm this scenario, we performed ARPES measurements along the $\Gamma-Z$ direction on the $x$ = 0.09 sample at various temperatures (Fig. 3(f-j)). The temperature-dependent ARPES spectra show that the positions of the $d_{xy}^-$ band remain unchanged over temperature. These results suggest that the band inversion is intact up to 175~K, preserving the topological invariant. We note that 175~K is well above the OSMP temperature at $x=0.19$~\cite{huang2022correlation}. Due to the spin-orbital separation from the Hund's coupling, the orbital-selective Mott transition-driven incoherence in the low energy spectrum emerges before the full suppression of the total spectral weight of the $d_{xy}$ orbital upon raising the temperature~\cite{hunter2023fate,sutter2019orbitally,kugler2022orbital}.

\begin{figure}[!t]
	\centering
	\includegraphics[width=0.5\textwidth]{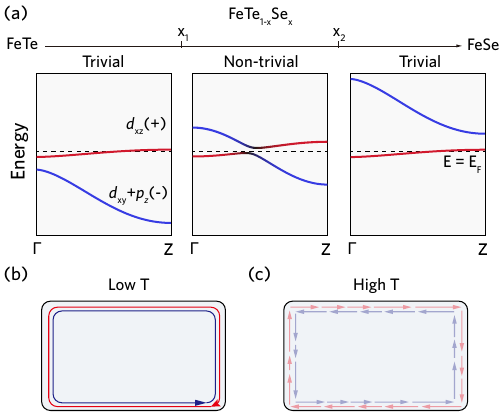}
	\caption{
         {\bf Correlation and topology.}
        (a) Schematic of band dispersion along $\Gamma-Z$. $x_1$ and $x_2$ denote the critical selenium contents in FTS where topological phase transitions occur.
        (b,c) Illustration of topological surface states at low and high temperatures.
	} 
	\label{Fig4}
\end{figure}

Given that lightly Se-doped FTS exhibits an orbital-selective Mott phase (OSMP) at high temperatures, we attribute the localized spin's fluctuation-driven modulation that scatters the topological surface state to the emergence of the OSMP. In FTS, the OSMP is characterized by the loss of coherence in the $d_{xy}$ orbital at high temperatures. The OSMP temperature for $x=0.19$ is approximately 50~K, which is consistent with the abrupt increase in linewidth shown in Fig.~3(e)~\cite{huang2022correlation}. Since the odd parity band responsible for the non-trivial topology mainly consists of $d_{xy}$ orbital, the onset of the OSMP significantly impacts the topological surface, diminishing its coherence and spectral weight in a manner analogous to its effect on the bulk bands. These aspects will be discussed in more detail in the Discussion section. It is also noteworthy that the pronounced broadening cannot be attributed to the phonon-driven thermal effect, as phonon-induced scattering is expected to be minimal at 50 K, considering the high Debye temperature and the robustness of the topological surface state against time-reversal-symmetric perturbations~\cite{maheshwari2018heat}.

\section{Discussion} \label{sec:discussion}

Our results provide crucial insights into how topological properties evolve under significant electron correlations, demonstrating that the OSMP affects topological properties via two distinct pathways. On one hand, correlation-driven band renormalization controls the band inversion condition, resulting in a topological phase transition. Specifically, we established a new lower critical doping $x_1$ between $x=0.04$ and $x=0.09$, expanding the phase diagram into a previously unexplored regime (Fig. 4(a)). On the other hand, our data reveals that OSMP-induced scattering critically limits the coherence of these surface states. Although the bulk topological invariant remains intact across the studied temperature range, the surface states become significantly broadened and weakened at elevated temperatures (Fig. 4(b-c)). We attribute this to strong scattering driven by spin fluctuations of the $d_{xy}$ orbital, which makes low-energy features indiscernible~\cite{yin2011kinetic}. This highlights a key distinction: while the system remains topologically non-trivial in a single-particle picture, the strong decoherence imposed by OSMP effectively deteriorates the practical robustness of the topological protection~\cite{hasan2010colloquium,rachel2018interacting}.

The robust $d_{xy}^-$ at elevated temperatures shown in Fig. 3(f-j) raises a question about the existence of truly zero quasiparticle residue (mass divergence) from the OSMP, in the presence of the inter-orbital hybridizations~\cite{kugler2022orbital}. As even FeTe shows robust $d_{xy}^-$~\cite{kim2023kondo}, the experimental realization of further localization of $d_{xy}$ beyond FeTe can give more clues on this question. Possible approaches are applying hydrostatic pressure~\cite{bendele2013pressure} or epitaxial strain on FeTe for further localization of $d_{xy}$~\cite{song2024growth,han2010superconductivity}.

Recently, we became aware of a similar report that also addressed the topology of FTS in the lightly Se-doped limit. In addition to Ref.~\cite{lin2025topological}, our discovery of the evolution of $d_{xy}^-$ along the $\Gamma-Z$ direction as a function of doping and temperature gives a deeper understanding of the underlying mechanism of OSMP and its relation to the topological phase transition.

\section*{Acknowledgements} \label{sec:acknowledgements}
We acknowledge S. Huh, Z.-X. Shen, D.-H. Lee, and G. Kotliar for useful discussions, and S. Jung and Y. Ishida for helpful comments on laser development. This work was supported by the National Research Foundation of Korea (NRF) grant funded by the Korean government (MSIT) (No. 2022R1A3B1077234) and Global Research Development Center Cooperative Hub Program (GRDC) through NRF (RS-2023-00258359). This work was also supported by the Institute of Applied Physics, Seoul National University. M.K. was supported by the National Research Foundation of Korea(NRF) grant funded by the Korea government(MSIT) (RS-2026-25478055), the Korea Institute for Advanced Study (KIAS) individual Grants (No. CG083502), Korea Basic Science Institute (National Research Facilities and Equipment
Center) Grant (No. RS-2024-00436672) funded by the Ministry of Education, Republic of Korea, and Korea Institute for Advancement of Technology (KIAT) grant funded by the Korea Government (MOTIE) (RS-2025-02214408, HRD Program for Industrial Innovation). The LQSGW+DMFT calculation is supported by the Center for Advanced Computation at KIAS. We acknowledge the MAX IV Laboratory for beamtime on the Bloch beamline under proposal 20241943. Research conducted at MAX IV, a Swedish national user facility, is supported by Vetenskapsrådet (Swedish Research Council, VR) under contract 2018-07152, Vinnova (Swedish Governmental Agency for Innovation Systems) under contract 2018-04969 and Formas under contract 2019-02496.

\section*{Author Contributions}
Y.K., M.K., and C.K. conceived the project. Y.K. and J.Y. synthesized and characterized FTS crystals. Y.K. developed the laser ARPES system with support from L.Y. Y.K. and S.K. conducted the laser ARPES experiments. Y.K., J.Y., and S.K. carried out the synchrotron ARPES experiments with support from K.T. and S.H. Y.K. analyzed the ARPES data. M.K. performed the calculations and theoretical analyses. The manuscript was written by Y.K., M.K., and C.K., with input and discussions from all authors.

\section*{Data availability}
The data that support the findings of this article are openly available at https://doi.org/10.5281/zenodo.16352003~\cite{kim_2025_16352003}.

\bibliography{TP.bib}

\clearpage
\onecolumngrid

\begin{center}
    \vspace*{1.0cm}
    \textbf{\Large Supplemental Material for \\ Fragility of topology under electronic correlations in iron chalcogenides} \\[1.5em]
    
    \normalsize
    Younsik Kim$^{1,*}$, Junseo Yoo$^1$, Sehoon Kim$^1$, Sungsoo Hahn$^2$, \\
    Kiyohisa Tanaka$^3$, Li Yu$^{4,5}$, Minjae Kim$^{6,7,\dagger}$, and Changyoung Kim$^{1,\ddagger}$ \\[1.0em]
    
    {\small \it
    $^1$Department of Physics and Astronomy, Seoul National University, Seoul, Korea \\
    $^2$MAX IV laboratory, Lund University, Lund, Sweden \\
    $^3$National Institutes of Natural Science, Institute for Molecular Science, Okazaki, Japan \\
    $^4$Beijing National Laboratory for Condensed Matter Physics and Institute of Physics, \\ Chinese Academy of Sciences, Beijing 100190, China \\
    $^5$University of Chinese Academy of Sciences, Beijing 100049, P. R. China \\
    $^6$Korea Institute for Advanced Study, Seoul, Korea \\
    $^7$Department of Semiconductor Science and Technology, Jeonbuk National University, Jeonju, Korea \\[1.0em]
    }
    
    {\small
    $^*$leblang@snu.ac.kr \\
    $^\dagger$mjkim1985@jbnu.ac.kr \\
    $^\ddagger$changyoung@snu.ac.kr
    }
    \vspace*{1.0cm}
\end{center}

------------------------------------
\setcounter{figure}{0}      
\setcounter{table}{0}       
\setcounter{equation}{0}   

\renewcommand{\thesection}{S\arabic{section}}
\renewcommand{\thefigure}{S\arabic{figure}}
\renewcommand{\thetable}{S\arabic{table}}
\renewcommand{\theequation}{S\arabic{equation}}

\newpage

\section*{1. Sample Characterization}
The actual stoichiometry of FeTe$_{1-x}$Se$_{x}$ was determined using energy-dispersive X-ray spectroscopy (EDX). Table~\ref{tab} presents the nominal compositions used during sample growth along with the measured (actual) compositions. We note that the EDX measurements were conducted after Te vapor annealing . The samples typically exhibit a lower Se ratio compared to the nominal composition, which may be attributed to the effects of the Te vapor annealing process.

\begin{table}[h]
\centering

\setlength{\tabcolsep}{10pt} 
\renewcommand{\arraystretch}{1.2} 

\begin{tabular}{cccccc}
\hline
\multicolumn{3}{c}{Nominal} & \multicolumn{3}{c}{Actual} \\
Fe & Te & Se & Fe & Te & Se \\
\hline
1   & 0.95 & 0.05 & 1.06 & 0.96 & 0.04 \\
1   & 0.90 & 0.10 & 1.05 & 0.91 & 0.09 \\
1   & 0.80 & 0.20 & 1.02 & 0.81 & 0.19 \\
1   & 0.67 & 0.33 & 1.01 & 0.69 & 0.31 \\
\hline
\end{tabular}
\caption{Nominal and actual sample compositions determined by energy-dispersive X-ray spectroscopy}
\label{tab}
\end{table}

Magnetic and superconducting properties were measured using a Quantum Design Magnetic Property Measurement System (MPMS). Figure S1 shows the magnetic moments obtained for samples with various selenium contents. The FeTe$_{1-x}$Se$_x$ sample with x = 0.04 exhibits a kink at 55 K, indicative of an antiferromagnetic transition. Samples with x = 0.09 and x = 0.19 display diamagnetic transitions at approximately 5 K and 13 K, respectively, consistent with superconducting transitions.

\begin{figure*}[!h]
	\centering
	\includegraphics[width=0.7\textwidth]{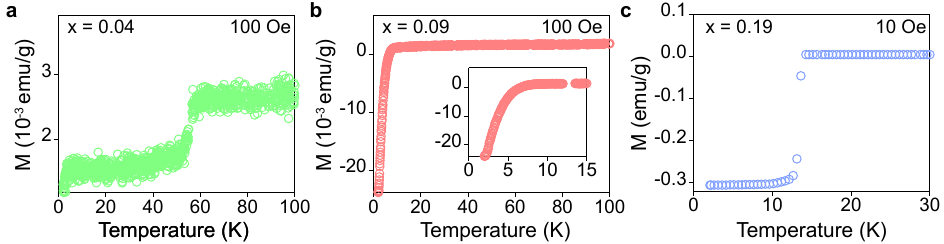}
	\caption{
         {\bf Temperature dependence of magnetization $M$ for FeTe$_{1-x}$Se$_x$ single crystals with varying Se content:}
         {\bf a} $x=0.04$,
         {\bf b} $x=0.09$,
	   {\bf c} $x=0.19$.
     Inset in {\bf b} highlights the superconducting transition at around 5~K.
	} 
	\label{Fig1}
\end{figure*}

\newpage
\section*{2. Experimental details} \label{sec:methods}

\textbf{Crystal growth and characterization}

FTS single crystals were grown by a modified Bridgman method~\cite{chen2009electronic,kim2023kondo}. A mixture of stoichiometric Fe granular (99.99~\%), Te pieces (99.999~\%), Se powder (99.99~\%) was sealed into an evacuated quartz tube and placed in a two-zone furnace. The temperature of the hot (cold) zone of the furnace was initially increased to 1070~$^{\circ}$C (970~$^{\circ}$C) with a rate of 100~$^{\circ}$C/h from room temperature and maintained 1070~$^{\circ}$C (970~$^{\circ}$C) for 12 hours. Subsequently, the temperature of the furnace was slowly decreased to 470~$^{\circ}$C (370~$^{\circ}$C) with a rate of 3~$^{\circ}$C/h for 200 hours. To remove excess iron, the grown crystals were annealed in Te atmosphere~\cite{lin2015role,xu2018coexistence, tevapor}. The as-grown crystals are cut into small pieces and placed in an evacuated quartz tube with additional tellurium at a molar ratio of 0.1. The furnace temperature is maintained at 400~$^{\circ}$C for 200 hours.

\textbf{Laser ARPES}

High-resolution, laser-based ARPES measurements were
performed using a lab-based system at Seoul National University. A home-built Yb-doped fiber laser with a center wavelength of 1064 nm, a pulse width of 50 ps and a repetition rate of 1.5 MHz was used to generate 7 eV (177 nm) light. Spectra were acquired using a Scienta ARTOF 10k electron analyzer with p-polarized light. Samples were cleaved \textit{in situ} along the (001) direction at 6~K in an ultra-high vacuum better than $6 \times 10^{-11}$ Torr. The energy resolution was set to better than 2 meV. The beam spot size of the 7 eV light is approximately $40~\mu m \times 30~\mu m$. A closed-loop sample position autocorrection system is applied to compensate for thermal drift during the temperature-dependent measurements~\cite{duan2022sample}.

\textbf{Synchrotron ARPES}

Synchrotron-based ARPES measurements were performed at beamline 7U at Ultraviolet Synchrotron Orbital Radiation (UVSOR) Facility, Institute for Molecular Science, Japan and beamline Bloch at MAX IV Laboratory. Spectra were acquired using A-1 analyzer from MB Scientific AB (UVSOR) and DA-30 from Scienta Omicron (MAX IV). Samples were cleaved \textit{in situ} at 15~K in an ultra-high vacuum better than $5 \times 10^{-11}$ Torr. The energy resolutions were set to better than 30~meV (UVSOR) and 15~meV (MAX IV). The $k_z$ dispersions shown in Fig. 2 and 3 were obtained with photon energy-dependent measurements, ranging from 10 eV to 28 eV.

\textbf{Computational Methods}

For the computation of the electronic structure of FeTe$_{1-x}$Se$_x$, we employed the linearized quasiparticle self-consistent GW plus dynamical mean-field theory (LQSGW+DMFT) method~\cite{kutepov2012electronic,kutepov2017linearized,georges1996dynamical,metzner1989correlated,muller1989correlated,brandt1989thermodynamics,janivs1991new,georges1992hubbard,jarrell1992hubbard,rozenberg1992mott,georges1992numerical,choi2016first,choi2019comdmft}, proven to capture the electronic structures of FeTe$_{1-x}$Se$_x$ with orbital selective correlations and topologically non-trivial phases~\cite{kim2024orbital}. To simulate the effects of the Te ratio variation, we vary the Se heights in the FeSe chemical composition, while fixing the lattice constant to the experimental value of FeTe$_{0.5}$Se$_{0.5}$~\cite{tegel2010crystal,li2009first}, which was shown to capture the Te ratio-driven topological phase transition of FeTe$_{1-x}$Se$_x$ linked to the orbital selective Mott transition~\cite{kim2024orbital}. The enhancement of the Te ratio is addressed by increasing the Se heights~\cite{kim2024orbital}. The Se heights of 1.48, 1.55, and 1.60 Angstrom correspond to the topologically non-trivial phase, the boundary for the topological phase transition, and the topologically trivial phase, respectively. For the computation of the band inversion driven hybridization gap and the topological surface band in the case of the topologically non-trivial phase, we included atomic spin-orbit coupling (SOC) to the LQSGW+DMFT electronic structure (LQSGW+DMFT+SOC) in the Fe(d) and the Se(p) orbitals, 42 meV and 450 meV, respectively, considering the correlation driven renormalization of the spin-orbit coupling for the Fe(d) orbital~\cite{kim2024orbital,kim2021spatial}. The temperatures are set as 300 K for the LQSGW+DMFT+SOC electronic structure for the topological surface states and 150 K for the LQSGW+DMFT electronic structure for the bulk electronic structure comparison between topologically trivial and non-trivial phases.

\newpage
\section*{3. $k_z$ determination}
$k_z$ of the synchrotron ARPES data shown in the main manuscript is calculated using an inner potential $V_0$ of 9.3 eV. The inner potential value is empirically determined using the periodicity of the ARPES data. The c-axis lattice constants of FTS are determined by using Vegard's law with the lattice constants of FeTe and FeSe as shown in Table~\ref{lattice}.

\begin{table}[h]
    \centering
    \setlength{\tabcolsep}{10pt} 
    \begin{tabular}{c c}
        \hline
        $x$ & $c$ (\text{\AA}) \\
        \hline
        0.04 & 6.27 \\
        0.09 & 6.23 \\
        0.19 & 6.15 \\
        0.31 & 6.06 \\
        \hline
    \end{tabular}
    \caption{\bf Out-of-plane lattice constants of FTS.}
    \label{lattice}
\end{table}

\newpage
\section*{4. Two-dimensional Fittings}

To accurately extract the dispersion of the $d_{xy}^-$ band, particularly near the band maximum where the spectral weight is cut by the Fermi edge, we employed a global two-dimensional (2D) fitting analysis~\cite{kurleto2021two}. The experimental ARPES spectra were modeled using a single-particle spectral function $A(E, \mathbf{k})$ with a finite quasiparticle lifetime, defined as:

\begin{equation}
f(E, \mathbf{k}) = \frac{\Gamma(E, \mathbf{k})/2}{\left[E - \varepsilon^{\star}(\mathbf{k})\right]^2 + \left[\Gamma(E, \mathbf{k})/2\right]^2}
\label{eq:spectral_function}
\end{equation}

Here, the imaginary part of the self-energy $\Gamma(E, \mathbf{k})$ and the bare band dispersion $\varepsilon^{\star}(\mathbf{k})$ are assumed to follow Fermi liquid-like dispersion: $\Gamma(E, \mathbf{k}) = \Gamma_0 + aE^2$ and $\varepsilon^{\star}(\mathbf{k}) = E_0 + b(k - k_0)^2$. Based on the equation, we first simulate data and compare it with the actual data. The optimal parameters can be found by minimizing the objective function $S$, defined as the residual sum of squares: 

\begin{equation}
S=\sum_{E,k}[A_{data}(E,k)-A_{sim}(E,k)]^2
\end{equation}

Here, $A_{data}$ and $A_{sim}$ refer to the obtained data and the simulated data, respectively. The simulated spectra were generated by multiplying this spectral function by the Fermi-Dirac distribution and convolving with the instrumental resolution, followed by normalization using integrated energy distribution curves.

We emphasize that this global 2D approach is crucial for our analysis. Standard one-dimensional (1D) momentum distribution curve (MDC) fitting becomes unreliable near the band top ($E \approx E_F$) because the two branches of the parabolic dispersion merge and the spectral intensity is cut off by the Fermi edge. In contrast, our 2D fitting utilizes the global spectral information, allowing for a robust determination of the band extrema even in the presence of noise and thermal broadening. The robustness of this method was verified against synthetic data with known parameters. This feature is directly demonstrated in the simulated data: Figure S2 shows simulated data with added noise. It can be seen that the 2D fitting method is more robust and accurate in predicting the band top position than MDC fitting, which tends to overestimate the position.

The optimization of the fitting was performed using the differential evolution algorithm, a global stochastic method designed to avoid local minima~\cite{storn1997differential}. The objective function was defined as the sum of squared errors between the experimental and simulated spectra. The error bars presented in the main text represent the standard deviation of the fitted parameters, estimated from an ensemble of fits with varied fitting ranges and initial seeds. Figure~\ref{Fig2} displays the ARPES data around the $\Gamma$ point along with the corresponding fitted results.

\begin{figure*}[!h]
	\centering
	\includegraphics[width=0.6\textwidth]{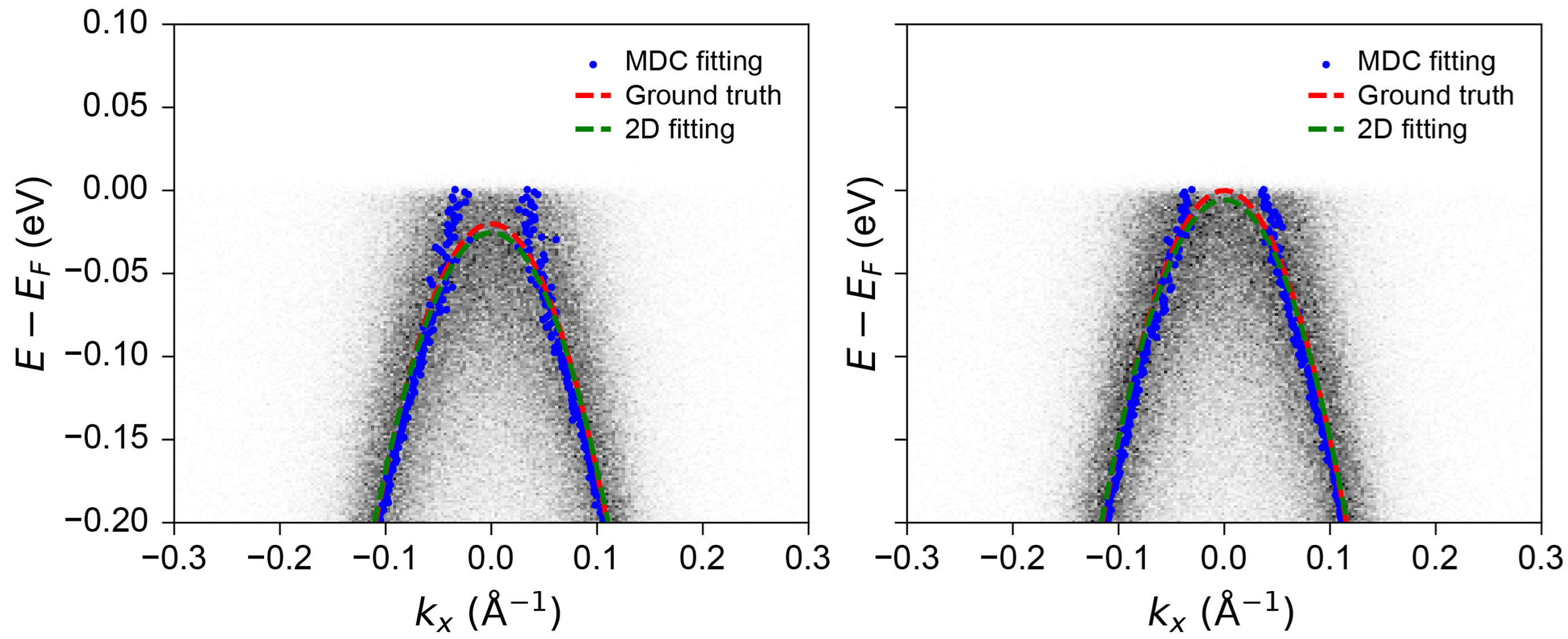}
	\caption{
         {\bf Demonstration of different fitting methods on simulated data. }
         The band top position is set to – 20 meV (left panel) and 0 meV (right panel).
	} 
	\label{2d}
\end{figure*}
\begin{figure*}[h]
	\centering
	\includegraphics[width=0.5\textwidth]{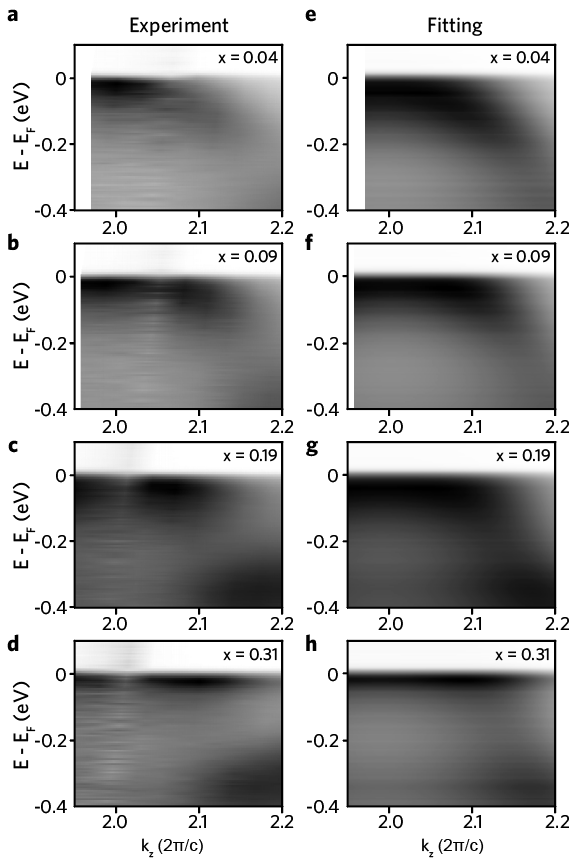}
	\caption{
         {\bf Two-dimensional fitting results}
         {\bf a-d} Experimental spectra obtained near the $\Gamma$ point with Se contents of 0.04, 0.09, 0.19, and 0.31, respectively.
        {\bf e-h} Corresponding two-dimensional fittings of {\bf a-d}.
	} 
	\label{Fig2}
\end{figure*}

\clearpage
\section*{5. Photon energy-dependent measurements at a different Brillouin zone}

\begin{figure*}[!h]
	\centering
	\includegraphics[width=0.7\textwidth]{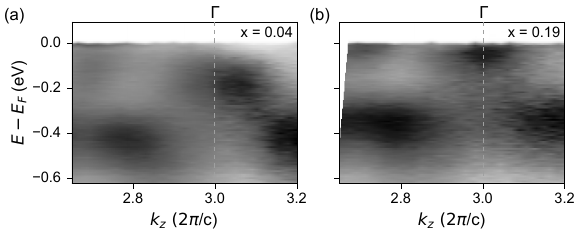}
	\caption{
         {\bf Band dispersion along $\Gamma-Z$ at a different Brillouin zone obtained with p-polarized light with selenium contents of (a) x = 0.04 and (b) x = 0.19.}
	} 
	\label{block_kz}
\end{figure*}

To further resolve the $d_{xy}^-$ band shown in the manuscript, we additionally conducted photon energy-dependent measurements at a different Brillouin zone. Fig.~\ref{block_kz} shows the band dispersion with $x=0.04$ and $x=0.19$. The $d_{xy}^-$ band does not cross the Fermi level for $x=0.04$, whereas it crosses the Fermi level for $x=0.19$. These results further support the topological phase transition scenario.

\newpage
\section*{6. Polarization-dependent measurement}

\begin{figure*}[!h]
	\centering
	\includegraphics[width=0.4\textwidth]{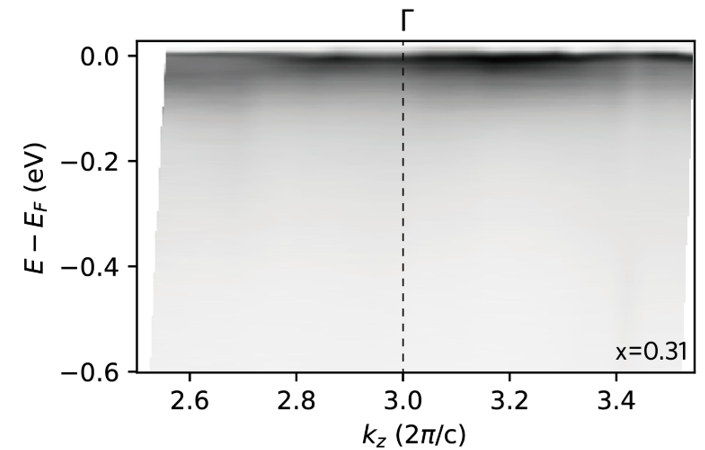}
	\caption{
         {\bf Band dispersion along $\Gamma-Z$ obtained with s-polarized light with selenium contents of $x = 0.19$.}
	}
	\label{spol}
\end{figure*}

The assignment of the $d_{xy}^-$ band is further supported by polarization-dependent measurement. Specifically, the dispersive band along $k_z$ is completely suppressed when measured using $s$-polarized light (Fig.~\ref{spol}). This strong polarization dependence is expected, as the $p_z$ orbital—the main contributor to the spectral weight—has a non-zero photoemission matrix element only for $p$-polarized light.

\newpage
\section*{7. Momentum distribution curve fittings}

\begin{figure*}[!h]
	\centering
	\includegraphics[width=0.3\textwidth]{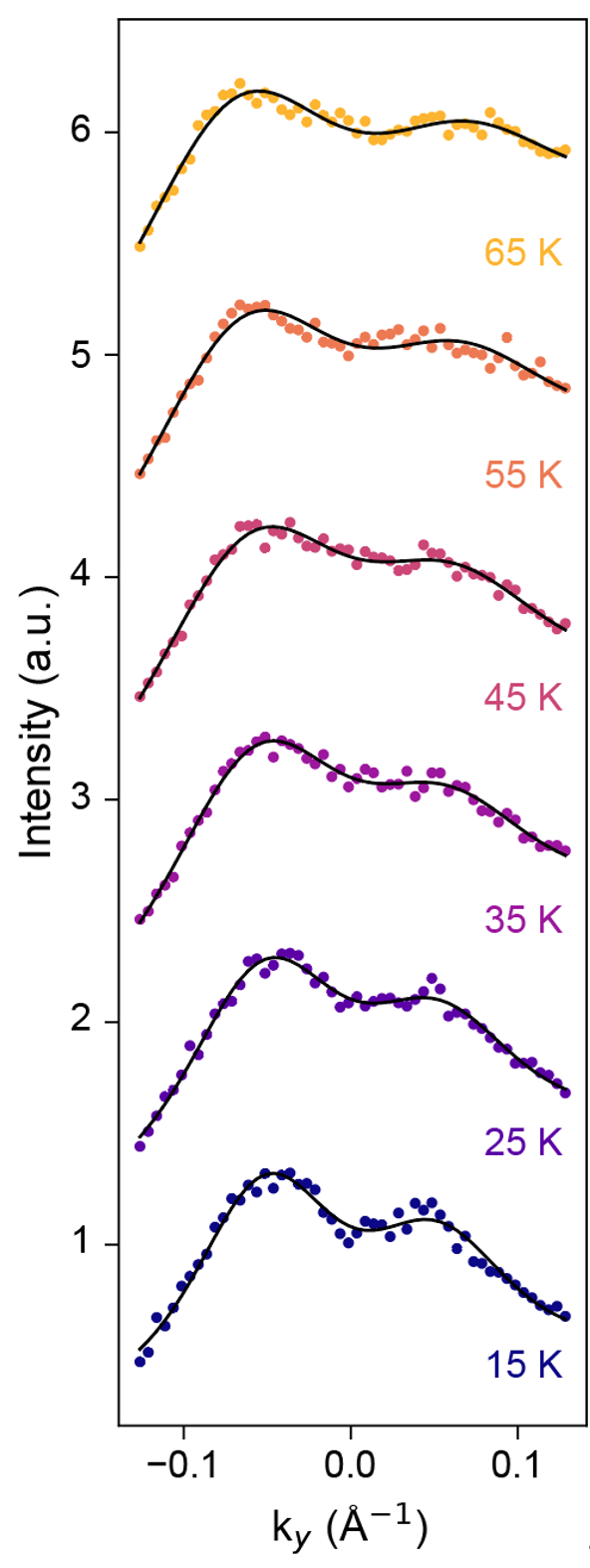}
	\caption{
         {\bf Temperature-dependent momentum distribution curves at the Fermi level for a selenium content of $x = 0.19$ at $k_x=0$.} The colored dots represent actual data and the black solid lines represent the fitted results.
	} 
	\label{mdc}
\end{figure*}

The data shown in Fig. 3(e) are extracted from the momentum distribution curve (MDC) fitting. MDCs are extracted at the Fermi level and are shown in Fig.~\ref{mdc}. The colored symbols represent the actual data, and the black line represents the fitted result. We employed a two-peak Lorentzian fitting. Here, the positions and widths of the two peaks were constrained to be the same.

\newpage
\section*{8. Experimental geometry and photoemission selection rule}

\begin{figure*}[!h]
	\centering
	\includegraphics[width=0.3\textwidth]{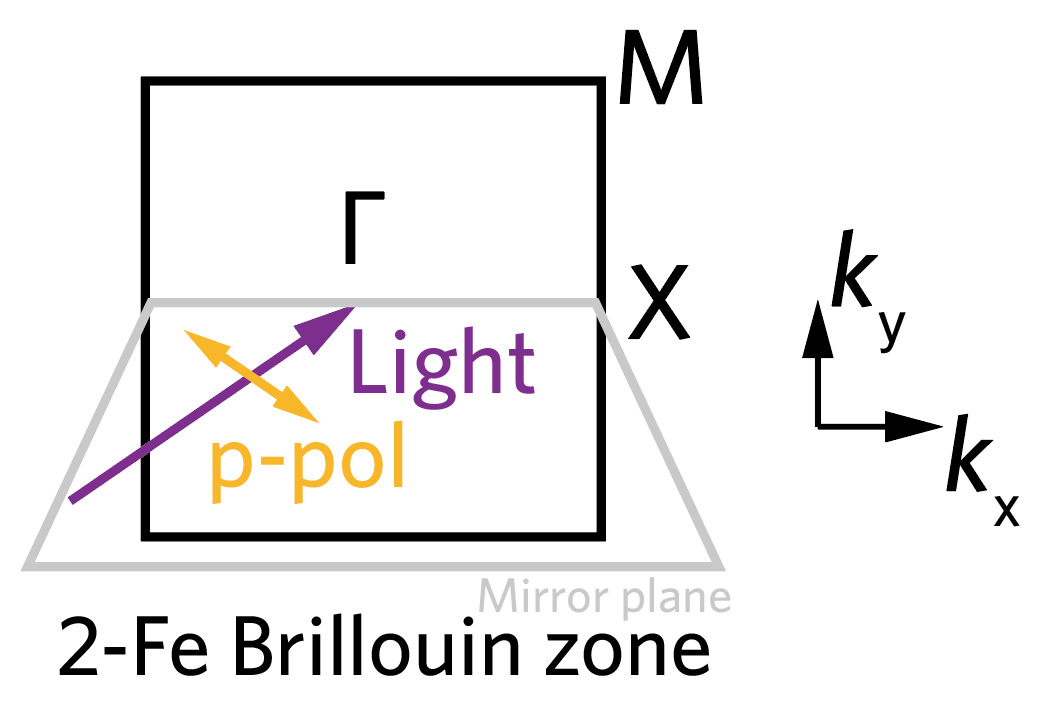}
	\caption{
         \bf Schematic of experimental geometry.
	} 
	\label{geo}
\end{figure*}

The experimental geometry for both laser- and synchrotron-based ARPES  is illustrated in Fig.~\ref{geo} unless otherwise noted. The sample is aligned along the $\Gamma - X$ direction and the light polarization is set to $p$-polarization. This experimental geometry is particularly sensitive to the $d_{xy}^-$ band, as the photoemission is allowed both from $d_{xy}$ and $p_z$ orbitals, which constitute the $d_{xy}^-$ band as summarized in Table S1~\cite{kim2023kondo,li2024orbital}.

\begin{table}[h]
    \centering
    \setlength{\tabcolsep}{10pt} 
    \begin{tabular}{c c c}
        \hline
        $\Gamma-X$ geometry & $p$-pol & $s$-pol\\
        \hline
        $d_{xy}$ orbital & O & X \\
        $p_{z}$ orbital & O & X \\
        \hline
    \end{tabular}
    \caption{\bf Photoemission selection rule for the $\Gamma–X$ geometry.}
    \label{selection}
\end{table}

\end{document}